\documentclass[10pt,conference]{IEEEtran}
\usepackage{etoolbox}
\newtoggle{devMode}
\usepackage[noadjust]{cite}

\togglefalse{devMode}

\usepackage[utf8]{inputenc}
\usepackage{url}
\usepackage{lipsum}
\usepackage{rotating}
\usepackage{booktabs}
\usepackage{amsmath}
\usepackage{colortbl}
\usepackage{array}
\usepackage{xspace}
\usepackage{hyperref}
\usepackage{xcolor}
\usepackage{comment}
\usepackage{listings}
\usepackage[font=small]{caption}
\usepackage{subcaption}
\usepackage{flushend}
\usepackage{makecell}

 \iftoggle{devMode}{
    \usepackage{todonotes}
 }{
    \usepackage[disable]{todonotes}
 }

\usepackage{tikz}
\usetikzlibrary{external}
\usepackage{pgfplots}
\usetikzlibrary{pgfplots.groupplots}
\usepackage{pgfplotstable}
\usetikzlibrary{arrows}
\usetikzlibrary{arrows.meta}
\usetikzlibrary{patterns}
\usetikzlibrary{positioning}
\usetikzlibrary{decorations.pathreplacing}
\usetikzlibrary{shapes.arrows}
\usetikzlibrary{shapes.geometric,shapes.misc}
\usetikzlibrary{pgfplots.groupplots}
\usetikzlibrary{
  circuits,
  circuits.logic.US
}
\pgfplotsset{compat=newest}
\usepackage{letltxmacro}
\reversemarginpar
\usepackage{todonotes}
\usepackage{marginnote} 

\LetLtxMacro{\oldtodo}{\todo}
\renewcommand{\todo}[2][]{\tikzexternaldisable\oldtodo[fancyline,size=\footnotesize,#1]{#2}\tikzexternalenable}
\renewcommand{\todo}[1]{\tikzexternaldisable\oldtodo[fancyline,size=\footnotesize]{#1}\tikzexternalenable}

\newcommand{\mypar}[1]{\medskip\noindent\textbf{#1}\xspace}
\newcommand{\ct}[1]{\textsc{\MakeLowercase{#1}}\xspace}

\newcommand{\basemagika}{Magika\xspace}
\newcommand{\magika}{\textsc{\basemagika}\xspace}

\newcommand{\eg}{e.g., }
\newcommand{\etc}{etc.}

\newcommand{\entitystyle}[1]{\texttt{#1}\xspace}
\newcommand{\file}{\entitystyle{file}}
\newcommand{\filemime}{\entitystyle{file-mime}}
\newcommand{\exif}{\entitystyle{exiftool}}
\newcommand{\trid}{\entitystyle{trid}}
\newcommand{\guesslang}{\entitystyle{guesslang}}
\newcommand{\libmagic}{\entitystyle{libmagic}}

\newcommand{\vscodeshort}{VS Code\xspace}

\newcommand{\cellmax}{\cellcolor[gray]{0.85}}

\def\ContentTypesVariantsNum{128\xspace}

\def\ContentTypesTextNum{43\xspace}
\def\ContentTypesBinaryNum{70\xspace}

\def\ContentTypesNum{113\xspace}  

\def\PerfEvalDatasetSamplesNum{1,130\xspace}

\def\DatasetTotalSamplesNumApprox{26.5 million\xspace}
\def\DatasetTotalSamplesNumApproxShort{26M\xspace}
\def\DatasetTrainSamplesNumApprox{24 million\xspace}
\def\DatasetTrainSamplesNumApproxShort{24M\xspace}
\def\DatasetValidationSamplesNumApprox{1.2 million\xspace}
\def\DatasetTestSamplesNumApprox{1.2 million\xspace}
\def\DatasetTestSamplesNumApproxShort{1.2M\xspace}

\title{\basemagika: AI-Powered Content-Type Detection}
\author{
\IEEEauthorblockN{\quad Yanick Fratantonio\quad}
\IEEEauthorblockA{Google}
\and
\IEEEauthorblockN{\quad Luca Invernizzi\quad}
\IEEEauthorblockA{Google}
\and
\IEEEauthorblockN{\quad Loua Farah\quad}
\IEEEauthorblockA{Google}
\and
\IEEEauthorblockN{\quad Kurt Thomas\quad}
\IEEEauthorblockA{Google}
\and
\IEEEauthorblockN{\quad Marina Zhang\quad}
\IEEEauthorblockA{Google}
\and
\IEEEauthorblockN{\quad\quad\quad\quad Ange Albertini\quad\quad\quad\quad}
\IEEEauthorblockA{Google}
\and
\IEEEauthorblockN{\quad\quad\quad\quad Francois Galilee\quad\quad\quad\quad}
\IEEEauthorblockA{Google}
\and
\IEEEauthorblockN{\quad\quad\quad\quad Giancarlo Metitieri\quad\quad\quad\quad}
\IEEEauthorblockA{Google}
\and
\IEEEauthorblockN{\quad\quad\quad Julien Cretin\quad\quad\quad}
\IEEEauthorblockA{Google}
\and
\IEEEauthorblockN{\quad\quad\quad Alex Petit-Bianco\quad\quad\quad}
\IEEEauthorblockA{Google}
\and
\IEEEauthorblockN{\quad\quad\quad David Tao\quad\quad\quad}
\IEEEauthorblockA{Google}
\and
\IEEEauthorblockN{\quad\quad\quad Elie Bursztein\quad\quad\quad}
\IEEEauthorblockA{Google}
}

\date{}

\begin{document}

\maketitle

\iftoggle{devMode}{
    \thispagestyle{plain}
    \pagestyle{plain}
 }{}
 
\begin{abstract}
The task of content-type detection---which entails identifying the data encoded in an arbitrary byte sequence---is critical for operating systems, development, reverse engineering environments, and a variety of security applications. In this paper, we introduce \magika, a novel AI-powered content-type detection tool. Under the hood, \magika employs a deep learning model that can execute on a single CPU with just 1MB of memory to store the model's weights. We show that \magika achieves an average F1 score of 99\% across over a hundred content types and a test set of more than 1M files, outperforming all existing content-type detection tools today.
In order to foster adoption and improvements, we open source \magika under an Apache 2 license on GitHub and will make our model and training pipeline publicly available. Our tool has already seen adoption by the Gmail email provider for attachment scanning, and it has been integrated with VirusTotal to aid with malware analysis.

We note that this paper discusses the first iteration of \magika, and a more recent version already supports more than 200 content types. The interested reader can see the latest development on the \magika GitHub repository, available at \href{https://github.com/google/magika}{github.com/google/magika}.

\end{abstract}

\section{Introduction}

Content-type detection is a fundamental computing task that identifies the data encoded in an arbitrary byte sequence. This allows an application to distinguish source code (C++, Python, \etc), media (PDF, JPG, \etc), binaries (EXE, ELF), and a variety of other file formats. As such, content-type detection impacts a wide range of downstream use-cases including software development tools, security tools, browsers, and media players. For example, development environments (e.g., Visual Studio Code, \vscodeshort for short) rely on content-type detection to decide which syntax highlighters and plugins to use. Security applications rely on content-type detection for policy enforcement (e.g., email providers prohibiting executable attachments), for forensic analysis and recovery, and for routing samples to the most capable content-type-specific threat analysis scanners (\eg VirusTotal and other anti-viruses have specialized scanners for binaries, scripts, or PDFs; these scanners are too resource-intensive to be run on all samples).

The need for content-type detection stems from byte sequences lacking an intrinsic, trustworthy indicator of their underlying file format. While some approaches to built-in indicators exist---such as file extensions, MIME types, magic bytes, and metadata---they can easily be omitted when a file's content is copied or transmitted, or otherwise spoofed (\eg to evade security systems that prohibit certain content types).

\begin{figure}[t]
    \centering
    \includegraphics[width=.98\columnwidth]{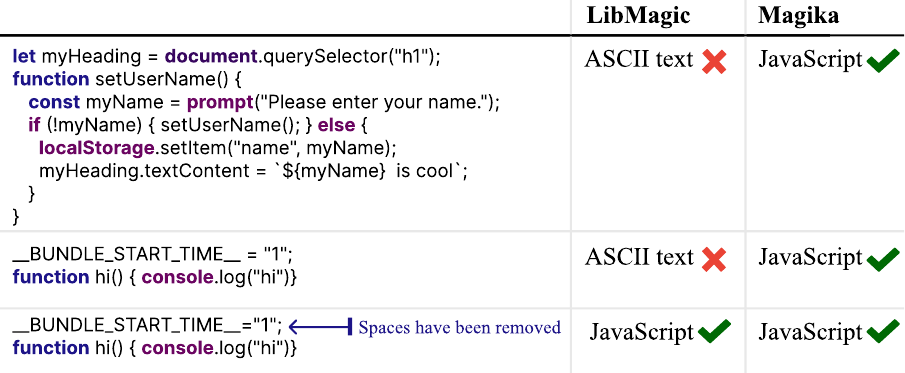}
    \caption{Example of the frailty of signature-based content-type detection when applied to distinct code snippets taken from  ``JavaScript Basics'' of the Mozilla's MDN~\cite{js-example}. \file---which relies on regular expressions for content-type detection---imprecisely labels each snippet as ASCII text unless the spaces around the ``='' sign are removed. As we show, our proposed content-type detector \magika overcomes these robustness limitations.}
    \label{fig:js-example}
\end{figure}

The task of content-type detection was first tackled with the \texttt{file} command in Bell Labs UNIX over five decades ago~\cite{file-man-1973}. Since then, developers have created a variety of tools, the vast majority of which rely on manually-written signatures (\eg rules or regular expressions) that are updated as new file types and versions emerge. Examples include the latest version of \file~\cite{file} (powered by \libmagic~\cite{libmagic-man}); \exif~\cite{exiftool}; and \trid~\cite{trid}. Unfortunately, signature-based approaches are frail: a single whitespace change (or other subtle byte sequence modification) can result in inaccurate content-type detection as shown in Figure~\ref{fig:js-example}. Beyond signatures, \guesslang~\cite{guesslang} employs a simple TensorFlow text model to distinguish source code content types, which is integrated into \vscodeshort~\cite{vscode-guesslang}.

In this paper, we discuss the design and implementation of a novel AI-powered content-type detection tool called \magika. Built using a deep learning model, \magika automatically infers the content type of a byte sequence without any reliance on human expertise with respect to the intricacies of different file formats. The model takes as input three sequences of 512 bytes drawn from the beginning, middle, and end of a file's content; automatically identifies patterns unique to content types; and outputs the most probable content type. We trained \magika to identify \ContentTypesNum canonical content types using  \DatasetTrainSamplesNumApproxShort file samples sourced from GitHub~\cite{github} and VirusTotal~\cite{virustotal}. We find that \magika achieves an average F1 score of 99\% on a holdout test dataset of \DatasetTestSamplesNumApproxShort samples.

For comparison, we benchmark \magika against four existing tools---\file, \exif, \trid, and \guesslang---to compare their accuracy, speed, and overall resource usage using the same test dataset of \DatasetTestSamplesNumApproxShort samples. We find that \magika outperforms every existing tool, with a 4\% F1 gain over the best tool for binary content types and a 22\%  F1 gain over the best tool for text content types, and 12\% F1 gain overall. The F1 gain increases, respectively, to 9\%, 47\% and 27\% when considering all content types in our benchmark, instead of just those supported by existing tools. For bulk inferences, \magika requires 5.77ms to yield a decision per file, outperforming all other existing tools except for \file (0.75ms). Likewise, \magika requires only 1MB of memory to represent model weights and achieves the aforementioned performance with just a single CPU; no GPU is required.

In order to foster wider adoption and accelerate improvements, we open source \magika under an Apache 2 license on GitHub~\cite{magika-github}. Upon release, we reached 4K stars on GitHub in less than a week; we currently total more than 7.7K stars.
In terms of real-world deployments, \magika has already been integrated within Gmail and Google Drive to predict the content type for hundreds of billions of files every week~\cite{magika-gmail-blogpost}. \magika has also been integrated with VirusTotal to assist with malware analysis, and we are in discussions to integrate \magika with \vscodeshort as a replacement for \guesslang. This positive reception reflects the real-world improvements \magika provides over existing software engineering toolsets.

\section{Related work}

Content-type detection is a well-studied, longstanding problem. Before presenting our new approach, we discuss both traditional approaches used by existing tools as well as research proposals to integrate machine learning into content-type detection.

\subsection{Traditional approaches to content-type detection}

The conventional approach to content-type detection relies on manually-crafted signatures (\eg regular expressions). 
The most prominent tool employing signatures is \file~\cite{file}, which is regarded as the default command-line utility for detecting content types.
\file provides support for binary file formats (which are particularly amenable to a signatures-based approach), including  highly-specialized formats (\eg file formats employed in specific video games for storing saved games, textures, etc.).
\file also possesses a range of signatures for textual content types (\eg C, Java, Python).
Anecdotal evidence suggests that \file exhibits higher accuracy for binary formats than textual ones, but its  performance has not been systematically evaluated prior to our study.

By default, \file outputs a textual description of the detected content type, and a number of additional metadata. For example, when processing a PDF file, \file will output its content type: PDF, but also the PDF version number, how many pages it has, and so forth.
Since this output is challenging to parse programmatically, \file also supports outputting a MIME type. MIME types are easier to parse and more codified than human descriptions, and thus, it may be more appropriate when integrating file in automated pipelines. However, dealing with MIME types has its set of challenges, which we document in Section~\ref{sec:existing}.

Other popular tools  include \trid~\cite{trid} and \exif~\cite{exiftool}. The latter was  originally  designed to  detect  image content types exclusively, but it has since expanded its scope. Apart from these general purpose content-type detection tools, there are a number of specialized tools that trade breadth for depth on a specific domain.
For instance, \texttt{PEiD}~\cite{peid} is a tool dedicated to detecting PE packers, whereas \texttt{Detect-It-Easy}~\cite{detect-it-easy} facilitates fine-grained inspection of PE files and other executable file formats. We omit these special-purpose tools from our comparative analysis due to their limited support of the content types we evaluate with \magika.

Lastly, numerous libraries re-implement a subset of existing features to offer content-type detection to various programming languages, often aiming to be free of dependencies to simplify integration. Examples include: \texttt{mime-types}~\cite{mime-types}, a library specializing in a limited number of content types, particularly for JavaScript clients; \texttt{PolyFile}~\cite{polyfile}, a pure-Python implementation of \texttt{libmagic}.  \texttt{filetype}~\cite{filetype}, a GoLang library for file type identification. \texttt{filetype.py}~\cite{filetype-py}, a Python library for file type identification; and \texttt{file-type}~\cite{file-type}, another (binary) file format detection tool for the JavaScript ecosystem. Given the derivative nature of many of these tools, we omit them from our comparative evaluation as well.

\subsection{Machine-learning approaches to content-type detection}

Recently, researchers have proposed a number of approaches that leverage machine learning to replace manual signatures. Sceadan~\cite{sceadan} proposes using support vector machines to model features based on unigram and bigram frequencies. Fitzgerald et al.~\cite{fitzgerald-nlp} suggest natural language processing techniques for file fragment classification, whereas Wang et al.~\cite{sparse-coding} propose using sparse coding and unsupervised learning to classify file fragments in the context of memory forensics.

More recent approaches employ appropriately state-of-the-art techniques such as recurrent and convolutional neural networks~\cite{fifty, bytercnn}.
These proposals have not found wide adoption, possibly due to their relatively low overall accuracy (ranging from 53\% to 84\%), limited support for different content types (between 18 and 75, primarily binary file formats), and competition from existing tools, which already provide adequate support for  binary file formats. As such, we treat them as out of scope for our comparative evaluation.

The most recent and successful approach is \guesslang~\cite{guesslang}, an open-source tool that detects programming languages from a snippet of source code.
\guesslang focuses exclusively on 54 textual content types and employs a Wide \& Deep TensorFlow model~\cite{cheng2016wide} for the detection task~\cite{guesslang-model}. Notably, \vscodeshort utilizes \guesslang  to infer the programming language when a user creates a new file without specifying a file extension~\cite{vscode-guesslang}.
While the original \guesslang is currently unmaintained (last commit in September 2021) and relies on deprecated TensorFlow abstractions, \vscodeshort developers maintain a Node.js client that facilitates the testing of the underlying model~\cite{vscode-language-detection}.

\section{Dataset}\label{sec:methodology-dataset}
We curate a novel dataset of \DatasetTotalSamplesNumApproxShort files drawn from a diverse set of \ContentTypesNum content types to act both as a benchmark for evaluating existing content-type detectors and to train our deep learning content-type detector. We describe the origin of this data, our validation steps, and limitations with our approach.

\mypar{Dataset sources.} In real-world settings, the distribution of content types varies from environment to environment. For example, social media uploads are dominated by videos and images, whereas digital-signing platforms predominantly receive PDFs. Rather than tailor our dataset to any single environment, we incorporate a sample of content types that may be present across a variety of environments including source code, executables, documents, media, archives, and more. We build a stratified dataset, in which every type has equal representation. When reporting performance metrics throughout this work, we break our results down per content type to account for this sampling strategy. In practice, any deployment environment can estimate the efficacy of \magika by taking a weighted average of our reported results, with weighting taking into account the frequency of content types within a specific setting.

To gather our stratified samples, we identified GitHub~\cite{github} (a popular source code development platform) and VirusTotal~\cite{virustotal} (a popular binary and file scanning platform) as two promising sources, as each covers a complementary set of content types: GitHub includes mostly text-based files such as source code (\eg C/C++, Java, Python), configuration files (\eg JSON, YAML, INI), and a variety of text files for documentation (\eg Markdown, RTF). Conversely, VirusTotal includes file archives (\eg ZIP, TAR), binaries (\eg EXE, DLL, ELF), as well as documents (\eg PDF, DOC, XLS).

\mypar{Content types.} 
We selected prevalent content types based on available public metrics. Using a public mirror of GitHub designed for large-scale analysis~\cite{github-bigquery-public-dataset} and VirusTotal's public API~\cite{virustotal-api}, we first calculate, for each, the top 50 most prominent file extensions across both platforms. We then identified the associated content type, if any, for each of these file extensions.\footnote{Not all file extensions are associated to a content type: for example, a significant portion of samples on VirusTotal have the \texttt{.file} or \texttt{.virus} extension which is not indicative of any content type.} Given the potential biases of our data sources, due to which potentially relevant content types do not appear in the most-frequent file extensions, we augment this list by consulting existing resources~\cite{wikipedia-list-file-formats,mozilla-list-file-formats,mimetype-io-all-types} and adding file extensions if at least one of the authors believed to be critical to make a first release of practical utility.

In total, we consider \ContentTypesVariantsNum file extensions in our study. As a simplification, we manually group some extensions into a single category. For example, we group both JPEG and JPG into a \ct{JPEG} content type. Similarly, we group both EXE and DLL into a \ct{PEBIN} content type. As a notation mechanic, we format these canonical content types as \ct{TYPE} throughout this paper, whereas we refer to file extensions as TYPE. This process reduces our final set of categories from \ContentTypesVariantsNum file extensions to \ContentTypesNum canonical content types. Of these, \ContentTypesBinaryNum are binary content types and \ContentTypesTextNum are text content types. Figure~\ref{fig:dataset-list-content-types} shows the full list of selected content types.

\begin{figure}[t]
    \centering
    \footnotesize
    
\begin{tabular}{c}
\toprule
\ct{ai}, \ct{apk}, \ct{appleplist}, \ct{asm}, \ct{asp}, \ct{batch}, \ct{bmp}, \ct{bzip}, \ct{c}, \ct{cab}, \\
\ct{cat}, \ct{chm}, \ct{coff}, \ct{crx}, \ct{cs}, \ct{css}, \ct{csv}, \ct{deb}, \ct{dex}, \ct{dmg}, \ct{doc},\\
\ct{docx}, \ct{elf}, \ct{emf}, \ct{eml}, \ct{epub}, \ct{flac}, \ct{gif}, \ct{go}, \ct{gzip}, \ct{hlp}, \ct{html},\\
\ct{ico}, \ct{ini}, \ct{internetshortcut}, \ct{iso}, \ct{jar}, \ct{java}, \ct{javabytecode}, \\
\ct{javascript}, \ct{jpeg}, \ct{json}, \ct{latex}, \ct{lisp}, \ct{lnk}, \ct{m3u}, \ct{macho}, \\
\ct{makefile}, \ct{markdown}, \ct{mht}, \ct{mp3}, \ct{mp4}, \ct{mscompress}, \ct{msi}, \\
\ct{mum}, \ct{odex}, \ct{odp}, \ct{ods}, \ct{odt}, \ct{ogg}, \ct{outlook}, \ct{pcap}, \ct{pdf},\\
\ct{pebin}, \ct{pem}, \ct{perl}, \ct{php}, \ct{png}, \ct{postscript}, \ct{powershell}, \ct{ppt},\\
\ct{pptx}, \ct{python}, \ct{pythonbytecode}, \ct{rar}, \ct{rdf}, \ct{rpm}, \ct{rst}, \ct{rtf},\\
\ct{ruby}, \ct{rust}, \ct{scala}, \ct{sevenzip}, \ct{shell}, \ct{smali}, \ct{sql}, \ct{squashfs},\\
\ct{svg}, \ct{swf}, \ct{symlinktext}, \ct{tar}, \ct{tga}, \ct{tiff}, \ct{torrent}, \ct{ttf}, \ct{txt},\\
\ct{unknown}, \ct{vba}, \ct{wav}, \ct{webm}, \ct{webp}, \ct{winregistry}, \ct{wmf}, \ct{xar},\\
\ct{xls}, \ct{xlsb}, \ct{xlsx}, \ct{xml}, \ct{xpi}, \ct{xz}, \ct{yaml}, \ct{zip}, \ct{zlibstream}\\

\bottomrule

\end{tabular}

    \medskip
    \caption{List of content types in our dataset.}
    \label{fig:dataset-list-content-types}

\end{figure}

We consider our selection of content types to be sufficiently large and diverse to demonstrate the generalizability of our approach. We leave the extension of our approach to even more content types to future work with the open source community.

\mypar{Sampling \& validating content types.} For each content type, we query GitHub and VirusTotal to obtain a random sample of files associated with the content type's file extensions. To minimize the risk of mislabeled content types in our ground truth, we perform a number of validation checks before adding a sample to our dataset. We specifically avoid using existing content-type detection tools as part of our validation, otherwise, we might oversimplify content-type detection (in the case of requiring agreement across all tools); or propagate detection errors (in the case of trusting one tool above others).

We use four heuristics for validation: file size, magic bytes (for binary files), character encoding (for text files), and file trustworthiness. For file size, we require any sample in our dataset to consist of at least 16 bytes. For magic bytes, we apply a set of \emph{necessary} but not \emph{sufficient} rules to validate a file extension. Note that these rules are, by design, straightforward, as we would otherwise risk the introduction of biases in our dataset. For example, all \ct{PEBIN} (the main Microsoft Windows executable format) \textit{must} start with the string \texttt{MZ} (\texttt{0x4D 0x5A}). Not all files starting with \texttt{MZ} are \ct{PEBIN} (e.g., they could be textual files which happened to start with the characters MZ), but if a sample's file extension claims to be an EXE or DLL, we verify this condition holds. For text files---where checks on magic bytes are not applicable---we 
merely verify the encoding to ensure the file contains only text characters. Finally, for samples from VirusTotal, we ensure that no anti-virus engines flags the sample as malicious due to such samples having an increased risk of obfuscated content types.

Independent of these workflows, we also create \emph{synthetic} samples for two content types: \ct{UNKNOWN} and \ct{txt}. For \ct{UNKNOWN}, each sample consists of a random byte sequence of arbitrary length. For \ct{txt}, each sample consists of a random string of text characters of arbitrary length. These synthetic samples are used as a form of data augmentation to train the model. We do so to help the model handle files with content types it has not been trained on, following best practices: rather than forcing the model to choose 1 of N valid content types, it can select \ct{unknown}.  We empirically observed that this approach has a negligible impact on accuracy on our dataset ($<$0.05\% accuracy improvement). Nevertheless, in adherence to best practices, we adopt this technique as it does not introduce any disadvantages.
The interested reader can find all the details about the entire dataset creation process in our open source release.

\mypar{Final dataset.} Our final dataset consists of \DatasetTotalSamplesNumApprox samples. We randomly split these into a training, validation, and testing dataset. The latter serves as a uniform benchmark for all content-type detection tools, including our deep learning model, but is never exposed to our model during training, nor has it been used to select the model hyperparameters or thresholds (for which we used the validation split, following best practices).  In total, we selected 10K samples per content type for our testing benchmark (with more than 1M samples in total, making our testing dataset large enough to obtain robust evaluation metrics), 10K samples for model validation, and 10K+ samples for model training (capping at 1 million per content type). There are two exceptions to this: for \ct{ISO} and \ct{ODP} content types, we have only 14K samples total. As we show later, this does not have a material difference on our deep learning detection accuracy. In total, our testing benchmark consists of \DatasetTestSamplesNumApprox samples; our training dataset consists of \DatasetTrainSamplesNumApprox samples; and our validation dataset consists of \DatasetValidationSamplesNumApprox samples.

\mypar{Limitations.} As with any measurement or machine learning study, our methodology incurs a number of limitations. First, our dataset consists of only \ContentTypesNum content types. We acknowledge the selected list is necessarily incomplete and may not encompass content types relevant to all deployment enviroments. While there is a long tail of many other content types in use today, acquiring and validating a representative sample is prohibitively expensive---in terms of manual overhead. As such, we focus on selecting a diverse and large enough number of content types to prove the generalizability of our deep learning approach on the most popular content types, leaving the extension of our technique to a more comprehensive set of content types to future work with the open source community. Second, despite our best efforts at validation and sampling, our benchmark dataset may contain mislabeled content types or be biased towards the samples available on VirusTotal or GitHub. As such, performance metrics may vary for other deployments, though our sample size should be suitably large to provide insights into the limitations of existing content-type detection tools, as well as our own.

\section{Benchmarking Existing Tools}
\label{sec:existing}

\begin{figure}[t]
    \centering
    \includegraphics[width=.98\columnwidth]{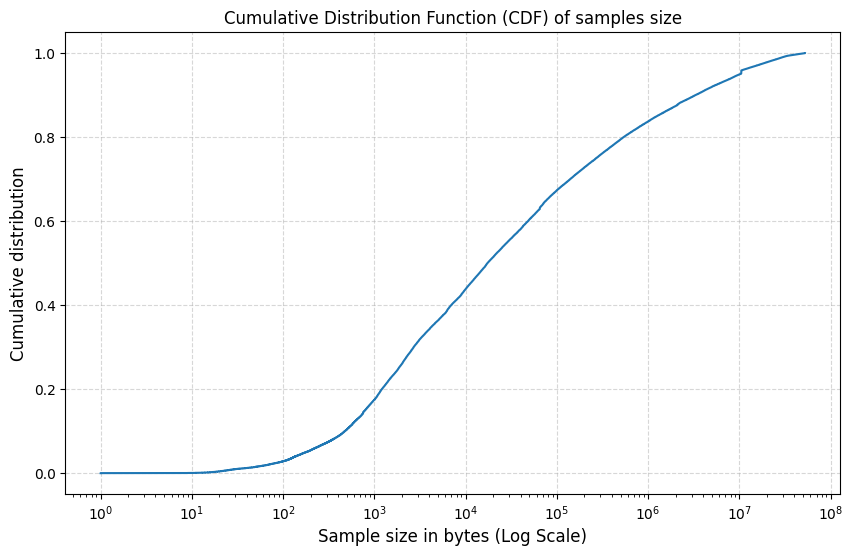}
    \caption{CDF of the sample sizes in our benchmark dataset.}
    \label{fig:samples-sizes-cdf}
\end{figure}

We benchmark the performance of existing content-type detection tools to motivate the need for more robust detection. Our measurements rely on the \DatasetTestSamplesNumApproxShort samples in our test benchmark dataset. The dataset includes samples of various sizes, as shown in Figure~\ref{fig:samples-sizes-cdf}. As part of our evaluation, we assess both the number of content types supported per tool (overlapping with the types in our benchmark) and the accuracy of predictions per tool. As we show, the best existing tool achieves an F1 score of only 88\%, with no single tool supporting all of the content types we evaluate.

\subsection{Tools Selection}
Our benchmark evaluates four popular content-type detection tools used for a variety of applications: 

\begin{itemize}
    \item \file~\cite{file}: the default command line tool for detecting content types for a variety of files. We note that \file can be optionally queried to return the MIME type, which we also benchmark as the two modes can yield distinct results (denoted \filemime).
    \item \exif~\cite{exiftool}: a tool originally developed for detecting image content types, but that has since expanded to a variety of binary and text content types.
    \item \trid~\cite{trid}: a tool for detecting a wide range of content types.
    \item \guesslang~\cite{guesslang}: a tool for detecting the programming language based on an excerpt of source code.
\end{itemize}
\vspace{5pt}

We select the first three because they represent the status-quo: they are based on well-established signatures-based approaches, readily maintained, and are widely adopted. We select \guesslang as a representative of emerging content type detection tools based on machine learning. \guesslang is unique as it focuses exclusively on textual content types, and is robust enough for real-world deployment in \vscodeshort~\cite{vscode-guesslang}.

\subsection{Metric selection}
For each tool, our benchmark assesses the precision and recall of inferred content types compared to the golden labels of our test dataset. Given this is a multi-class classification problem, we estimate per-type \textit{precision} as: $\text{TP} / (\text{TP} + \text{FP})$.
Here, a true positive (TP) indicates the tool predicts the correct golden label,\footnote{Given multiple fine-grained content types may be associated with a canonical content type (\eg JavaScript and TypeScript $\rightarrow$ \ct{JavaScript}), we treat a prediction as accurate if it matches any fine-grained content types in our mapping.} while a false positive (FP) indicates the tool predicts a specific content type, but in fact the golden label is any other content type. We calculate per-type \textit{recall} as: $\text{TP} / (\text{TP} + \text{FN})$.
Here, a false negative (FN) means a tool failed to detect a sample as the golden canonical content type, predicting any other label. For ease of presentation, we simplify these metrics into an F1 score, which provides equal weight to both per-type precision and recall:

\begin{equation*}
\footnotesize
F_1 = \frac{2 \cdot \text{Precision} \cdot \text{Recall}}{\text{Precision} + \text{Recall}}\vspace{5pt}
\end{equation*}

{\noindent}We calculate this F1 score per content type and then average F1 scores across text content types, binary content types, and overall content types to provide a variety of assessments on how well existing tools perform.

\subsection{Automating Large-Scale Evaluations}
The text string outputs of existing content type detection tools are not designed for cross comparison. As such, we develop a harness to programmatically query millions of samples while normalizing the outputs to match one of the \ContentTypesNum canonical content types in our benchmark.

\mypar{Normalizing detected content types.} Unfortunately, there are no canonical naming conventions for content types across tools---or even the same version of the same tool. Nor is there a hierarchy of how content types relate. For example, valid command line outputs of \file for XML documents include \texttt{XML document}, \texttt{XML 1.0 document}, and \texttt{XML 1.0 document text}, among others.
We also find that different versions of the same tool will silently update naming conventions. For example, \file silently changed the output for JavaScript files from ``Node.js script text executable'' to ``Node.js script executable,'' dropping the ``text'' keyword, and potentially breaking automated workflows~\cite{file-javascript-output-update}.

Similarly, we find that the MIME types generated by tools, despite being machine-oriented, also do not follow canonical naming conventions, for several reasons: MIME types change over time (e.g., \texttt{text/x-markdown} vs. \texttt{text/markdown}~\cite{markdown-mime-type-registration}); tools (silently) update their output (e.g., for \ct{PEBIN}, \file \texttt{5.41} outputs 
\texttt{app/x-dosexec}, while \file \texttt{5.44} outputs \texttt{app/vnd.microsoft.portable-executable}); some content types have multiple valid MIME types (e.g., for \ct{XML}, both \texttt{application/xml} and \texttt{text/xml} are valid); tools databases have typos (e.g., \texttt{text/pyton}), or arbitrary variations (e.g., \texttt{text/python37} vs. \texttt{text/python}).

We used a manual, iterative process to address all naming discrepancies. For each tool, we gathered the default command line outputs for every file in our benchmark. We then grouped these by output frequency, mapped the most popular types to one of our \ContentTypesNum canonical content types using a set of hand-written rules, and iteratively updated the set of rules until 99\% of the default outputs mapped to a canonical type. In the case a tool produced a content type outside our canonical set, we flag it with a special error code.

\mypar{Enumerating supported content types.} Of the tools we evaluate, only \exif and \guesslang provide a list of supported content types.
To determine which tool supports which content type, we use the normalized outputs over every sample as previously discussed. If a tool never flags any sample as one of our canonical content types, we make a simplifying assumption that it does not support that content type. This provides a fairer comparison with existing tools: we omit unsupported samples from some of our metrics, rather than giving a tool an F1 score of 0\% for such content types.
Table~\ref{tab:supported-content-types-existing-tools-summary} shows a breakdown of the content types in our benchmark supported by existing tools across text, binaries, and overall.

\begin{table}[t]
    \centering
    
\begin{tabular}{l|c|c|c}
    \textbf{Tool name} & 
    \textbf{\# Binary types} &
    \textbf{\# Text types} &
    \textbf{\# All types} \\
    & (Of 70) & (Of 34) & (Of 113) \\
\toprule

\textbf{\file} & 66 & \cellmax 31 & \cellmax 97 \\ \hline
\textbf{\filemime} & 64 & 27 & 91 \\ \hline
\textbf{\exif} & 38 & 14 & 52 \\ \hline
\textbf{\trid} & \cellmax 68 & 23 & 91 \\ \hline
\textbf{\guesslang} & 0 & 29 & 29 \\

\bottomrule

\end{tabular}

    \medskip
    \caption{Number of content types in our benchmark that are supported by existing content-type detection tools.}
    \label{tab:supported-content-types-existing-tools-summary}
\end{table}

\subsection{Results}
We report a high-level summary of our benchmark results for each existing tool in Table~\ref{tab:accuracy-eval-f1score-existing-tools-summary}. For space reasons, we share per-content type results for only a sample of the \ContentTypesNum content types we benchmark in Table~\ref{tab:accuracy-eval-f1score-existing-tools}. Full results can be found at~\cite{magika-anon-website}. Overall, if we restrict our view to content types supported by each tool, we find that \filemime achieves the best performance with an average F1 score of 88\%---with \exif and \trid scoring just 1\% lower. For text exclusively, \guesslang achieves the best performance with an average F1 score of 77\%. For binaries exclusively, \filemime achieves the best performance with an average F1 score of 96\%. As we demonstrate shortly, \magika is able to outperform all of these tools, achieving an overall average F1 score of 99\%. We explore the performance of each existing tool in detail below.

\begin{table}[t]
    \centering
    \footnotesize
    
\begin{tabular}{p{3.1cm}|c|c|c|c|c}
    \textbf{Content Type Metric} & 
    \rotatebox{90}{\textbf{\file}} &
    \rotatebox{90}{\textbf{\filemime}} &
    \rotatebox{90}{\textbf{\exif}} & 
    \rotatebox{90}{\textbf{\trid}} &
    \rotatebox{90}{\textbf{\guesslang}} \\
\toprule

Binary -- supported types & 92\% & \cellmax 96\% & 93\% & 93\% & - \\ \hline
Text -- supported types & 67\% & 68\% & 70\% & 68\% & \cellmax 77\% \\ \hline
Overall -- supported types & 84\% & \cellmax 88\% & 87\% & 87\% & 77\% \\
\\
\toprule
Binary -- all types & 87\% & 87\% & 50\% & \cellmax 91\% & 0\% \\ \hline
Text -- all types & 48\% & 42\% & 22\% & 35\% & \cellmax 52\% \\ \hline
Overall -- all types & \cellmax 72\% & 70\% & 39\% & 70\% & 19\% \\
\bottomrule

\end{tabular}

    \medskip
    \caption{Performance of existing content-type detection tools---measured as an F1 score---averaged across all binary, text, and overall content types. For fairness, we include two scopes for our metrics: performance restricted to supported content types, and performance on all content types.}
    \label{tab:accuracy-eval-f1score-existing-tools-summary}

\end{table}

\mypar{\file and \filemime.} We find that \file is highly performant for supported binary content types, with an average F1 score of 96\% when using the MIME flag, and 92\% otherwise. However, its accuracy decreases for some specific binary types, such as \ct{APK} (80\%), \ct{JAR} (64\%), or \ct{DMG} (4\%). For supported text types, the average F1 score drops to 68\%, likely due to the limitations of signatures when applied to text. We find the accuracy of supported text content types can vary widely, with an F1 score of 88\% for \ct{Ruby}, 70\% for \ct{Lisp}, and just 1\% for \ct{powershell}. Support is also missing for some programming languages, such as \ct{rust} and \ct{scala}. We note that the MIME flag is critical for some text contexts, with accuracy dropping from 100\% to 1\% for \ct{DOC} without the flag.

\begin{table}[t]
    \centering
    \footnotesize
    
\begin{tabular}{p{2cm}|c|c|c|c|c}
    \textbf{Content Type} & 
    \rotatebox{90}{\textbf{\file}} &
    \rotatebox{90}{\textbf{\filemime}} &
    \rotatebox{90}{\textbf{\exif}} & 
    \rotatebox{90}{\textbf{\trid}} &
    \rotatebox{90}{\textbf{\guesslang}} \\
\toprule

\ct{apk} & 80\% & 80\% & - & \cellmax 83\% & - \\ \hline
\ct{asm} & 21\% & 21\% & - & - & \cellmax 90\% \\ \hline
\ct{asp} & - & - & - & \cellmax 80\% & - \\ \hline
\ct{batch} & \cellmax 65\% & \cellmax 65\% & - & - & 38\% \\ \hline
\ct{bmp} & \cellmax 100\% & \cellmax 100\% & \cellmax 100\% & 98\% & - \\ \hline
\ct{c} & 59\% & 59\% & - & - & \cellmax 91\% \\ \hline
\ct{cs} & - & - & - & - & \cellmax 96\% \\ \hline
\ct{css} & - & - & - & - & \cellmax 84\% \\ \hline
\ct{csv} & \cellmax 75\% & \cellmax 75\% & - & - & 56\% \\ \hline
\ct{dmg} & 0\% & 4\% & - & \cellmax 77\% & - \\ \hline
\ct{doc} & 1\% & \cellmax 100\% & 99\% & 98\% & - \\ \hline
\ct{docx} & 99\% & 99\% & 65\% & \cellmax 100\% & - \\ \hline
\ct{elf} & \cellmax 100\% & \cellmax 100\% & - & 63\% & - \\ \hline
\ct{go} & - & - & - & - & \cellmax 95\% \\ \hline
\ct{html} & 58\% & 58\% & \cellmax 81\% & 73\% & 60\% \\ \hline
\ct{ini} & 2\% & 2\% & - & \cellmax 46\% & 28\% \\ \hline
\ct{jar} & 64\% & 64\% & - & \cellmax 74\% & - \\ \hline
\ct{java} & 81\% & 81\% & - & - & \cellmax 87\% \\ \hline
\ct{javascript} & 81\% & 81\% & - & - & \cellmax 88\% \\ \hline
\ct{jpeg} & \cellmax 100\% & \cellmax 100\% & \cellmax 100\% & \cellmax 100\% & - \\ \hline
\ct{json} & \cellmax 98\% & \cellmax 98\% & \cellmax 98\% & 9\% & 84\% \\ \hline
\ct{macho} & \cellmax 100\% & \cellmax 100\% & - & \cellmax 100\% & - \\ \hline
\ct{makefile} & 62\% & 62\% & - & - & \cellmax 95\% \\ \hline
\ct{markdown} & - & - & - & - & \cellmax 48\% \\ \hline
\ct{pdf} & \cellmax 100\% & \cellmax 100\% & \cellmax 100\% & \cellmax 100\% & - \\ \hline
\ct{pebin} & \cellmax 100\% & \cellmax 100\% & - & 56\% & - \\ \hline
\ct{perl} & 74\% & 75\% & 75\% & 3\% & \cellmax 80\% \\ \hline
\ct{png} & \cellmax 100\% & \cellmax 100\% & \cellmax 100\% & \cellmax 100\% & - \\ \hline
\ct{powershehll} & 1\% & - & - & - & \cellmax 91\% \\ \hline
\ct{python} & \cellmax 94\% & 87\% & 22\% & - & 90\% \\ \hline
\ct{ruby} & 88\% & 88\% & 3\% & - & \cellmax 89\% \\ \hline
\ct{shell} & \cellmax 89\% & \cellmax 89\% & 88\% & 61\% & 83\% \\ \hline
\ct{sql} & - & - & - & 6\% & \cellmax 82\% \\ \hline
\ct{txt} & \cellmax 17\% & 15\% & 10\% & 1\% & 0\% \\ \hline
\ct{unknown} & \cellmax 75\% & 31\% & 5\% & 6\% & 2\% \\ \hline
\ct{vba} & - & - & - & - & \cellmax 51\% \\ \hline
\ct{xls} & 12\% & \cellmax 99\% & 90\% & 91\% & - \\ \hline
\ct{xlsx} & \cellmax 100\% & \cellmax 100\% & 97\% & 99\% & - \\ \hline
\ct{xml} & 31\% & 31\% & 35\% & \cellmax 64\% & 31\% \\ \hline
\ct{yaml} & - & - & - & 0\% & \cellmax 74\% \\ \hline
\ct{zip} & 56\% & 56\% & 39\% & \cellmax 77\% & - \\

\bottomrule
\end{tabular}

    \medskip
    \caption{Performance of existing content-type detection tools---measured as an F1 score---per content type for a sample of content types in our benchmark.}
    \label{tab:accuracy-eval-f1score-existing-tools}
\end{table}

\mypar{\exif and \trid.}  We find that \trid and \exif provide similar overall accuracy, with an average F1 score of 87\% each. However, \trid supports far more content types (91) than \exif (52). Both \trid and \exif see drops in accuracy for text content types, with average F1 scores falling to 70\% and 68\% respectively. For instance, \trid produces inaccurate inferences for \ct{JSON} (9\%), \ct{SQL} (6\%), and \ct{YAML} ($<$1\%) despite supporting each content type. Likewise, \exif struggles with \ct{python} (22\%) and \ct{ruby} (3\%) among other text content types. As with \file, the high variance of F1 scores makes it challenging for clients to understand the quality of content-type detection from a tool absent benchmarks.

\mypar{\guesslang.} As previously mentioned, \guesslang focuses exclusively on text content types, achieving an average F1 score of 77\% for supported types. While \guesslang is highly accurate for some programming languages like \ct{Rust} (97\%) or \ct{Scala} (92\%), it nevertheless struggles with \ct{html} (60\%) and \ct{xml} (31\%) among others. One limitation specific to \guesslang is that it does not support an ``unknown'' verdict. As such, it still predicts a random, incorrect content type for our synthetic \ct{txt} samples. We note that \vscodeshort developers ran into this exact problem, as discussed in a GitHub issue~\cite{vscode-guesslang-problems}. Instead, \vscodeshort implements a number of heuristics to artificially penalize automatic recognition of content types prone to false positives~\cite{vscode-guesslang-heuristics}.

\section{\magika}
\label{sec:magika}

In light of the limitations of existing content-type detectors, we design and train a new deep learning classifier called \magika that distinguishes between content types without the need for signatures created by experts. 

\subsection{Requirements}
We design \magika to support two distinct deployment scenarios with a single model:  1) a command line tool that can replace established utilities, and 2) a bulk inference tool that can scale to analyzing billions of samples per day. These scenarios constrain how we approach accuracy, speed, and resource utilization.

\mypar{Accuracy.} \magika should provide equivalent or better accuracy at distinguishing between the \ContentTypesNum content types in our dataset. Furthermore, detection should depend exclusively on a file's content---regardless of the presence of a file extension or metadata. As before, we use an F1 score to evaluate accuracy.

\mypar{Speed.} Bulk inference should return a decision within $<$10ms per file, excluding any initialization overhead for loading a model into memory. Command line users may be sensitive to such initialization costs. As such, the total overhead of initialization and inference should should remain $<$100ms. This speed also includes the time spent reading a file, which has design implications for how best to detect the content type of large files (e.g., $>$100MB) without necessarily reading the whole file.

\mypar{Resources.} In order to maximize potential deployment scenarios, \magika should meet our speed and accuracy requirements even on a single CPU. This resource constraint is atypical of machine learning, which often assumes access to a GPU. However, we cannot expect command line users to have access to a GPU. Likewise, requiring large-scale platforms to dedicate GPU resources for the simple task of content type detection would be prohibitively expensive, especially considering the required scale (\eg potentially billions of files processed daily).

\subsection{Model Architecture}

\begin{figure}[t]
    \centering
    \includegraphics[angle=90,origin=c,width=.9\linewidth]{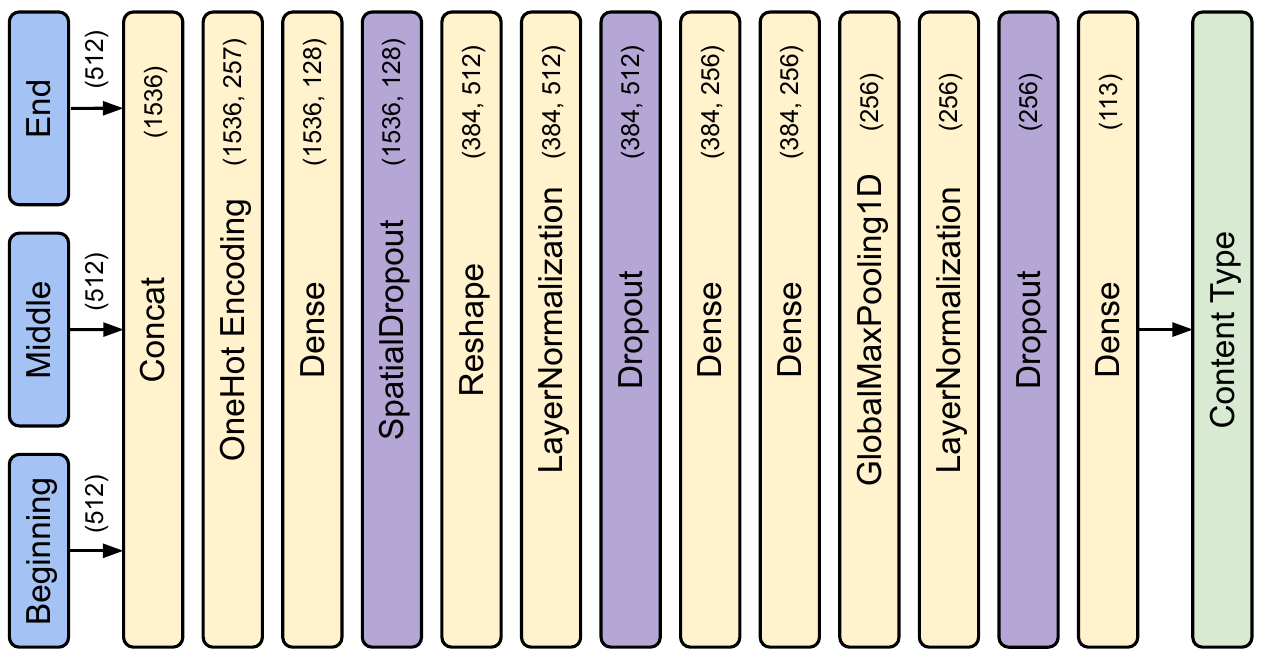}
    \vspace{-1.7cm}
    \caption{Architecture of \magika. The input and the output are depicted in blue and green, respectively. The model's layers are in yellow. The layers in purple are used only in training. The numbers next to the layers' names indicate the size of their outputs.}
    \label{fig:model-diagram}
\end{figure}

We present the architecture of the deep learning model that powers \magika in Figure~\ref{fig:model-diagram}. The overall flow consists of (1) transforming the contents of a file into a fixed-sized vector representation; (2) processing the vectors via a neural network; and (3) interpreting the model outputs to predict the most probable content type.

\mypar{Inputs.} The model's inputs consists of three vectors that encode a sequence of 512 bytes selected from the beginning, middle (taken from the center of the file, with no randomness), and end of an input file. We encode each byte as an integer in the range [0, 255]. To maximize the utility of the inputs made available to the model, we strip any leading or trailing whitespace from the beginning and end of a file's contents before selecting bytes for encoding. As the model's input has constant size (i.e., 3x512 integers), if an input file is too small, we pad the encoding with a special character (represented by the integer 256). We then concatenate the three input vectors to form a single vector of 3x512 integers, which we one-hot encode and embed into a 128-float vector using a Dense layer.

While we considered using longer sequences, or more sequences of bytes from a file, in practice this incurs additional resource requirements and slows down processing due to having to seek over more portions of the input file (e.g., files $>$ 100MB). By using a fixed size input rather than a whole file, we ensure inference is constant time. Despite limiting the model's visibility, we demonstrate we can still achieve robust accuracy.

\mypar{Trunk.} The model trunk consists of two components. First, the output of the previous Dense layer is reshaped to 384x512 dimensions, effectively reorganizing the previous output into small, four-byte chunks. Each chunk is represented by a 512 dimension vector that the model can treat as a single entity instead of having to consider each byte separately, which we found to improve both performance and efficiency. Then, the model contains two 256-dimensional Dense layers with \textit{gelu} activation~\cite{gelu-paper}. A global max pooling layer~\cite{globalmaxpool-keras} performs downsampling and reduces the dimensionality back to 1D. During training, we apply layer normalization~\cite{layernorm-keras}, a 10\% dropout rate~\cite{dropout-keras}, and a 10\% of spatial dropout rate~\cite{spatialdropout-keras} for regularization throughout the model.

This final model represents hundreds of design iterations across architectures that are simple enough to be small and fast on a CPU. We also ran extensive hyperparameter tuning experiments (using the validation dataset) and tested various model configurations that evaluated the embedding dimension; the size of the Reshape layer; the number, size, and activation of the Dense layers; the normalization type; and the amount of dropout applied before converging on this specific design.

\mypar{Outputs.}
The final Dense layer uses \textit{softmax} activation~\cite{softmax-paper} with size equal to the number of canonical content types in our dataset. The output vector represents a probability distribution over each of the potential content types. We can select the most likely output with an \textit{argmax} function~\cite{argmax-tf}, or apply individual confidence thresholds per content type, ultimately yielding the single most likely content type among those that meet a minimum confidence threshold.

\subsection{Training}

\begin{figure}[t]
    \includegraphics[width=\linewidth]{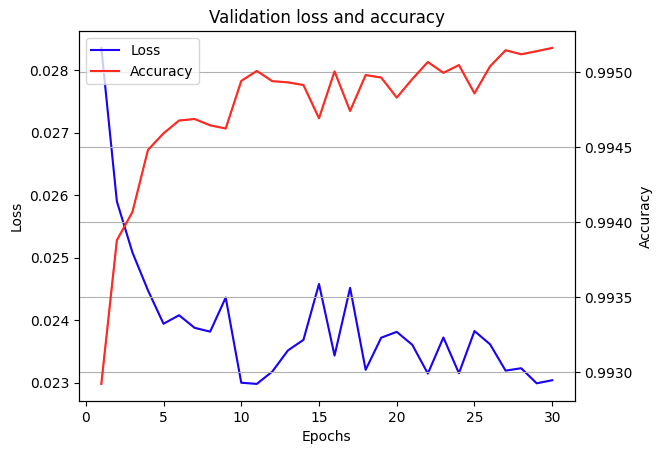}
    \caption{Validation loss and validation accuracy as the training progresses in terms of the number of epochs. We find accuracy increases up to around 30 epochs.}
    \label{fig:training-loss-val}
\end{figure}

\begin{figure}[t]
    \includegraphics[width=.9\linewidth]{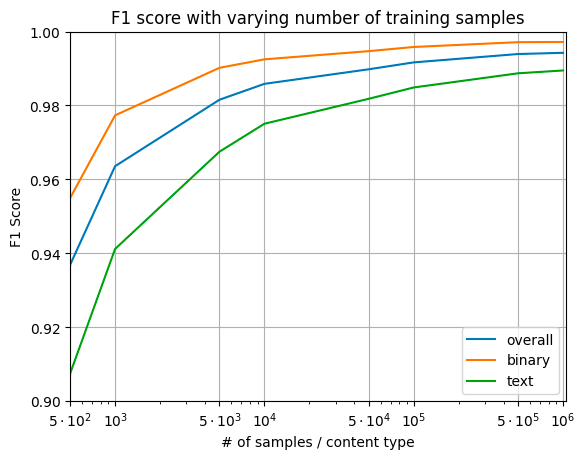}
    \caption{Average F1 Score (after one epoch of training) with increasing number of samples per content type, across binary, text, overall content types.}
    \label{fig:samples-num-vs-f1score}
\end{figure}

We trained the \magika model using Categorical Cross-Entropy loss which is suited for multi-class classification settings, with Adam as the optimizer (batch size $=256$, learning rate $=0.001$, $\beta_1=0.9$, $\beta_2=0.999$, and $\epsilon=1e-07$). Additionally, we use CutMix~\cite{cutmix} for data augmentation, although our experiments did not highlight a significant boost in validation accuracy. However, since CutMix is used only during training and does not affect model inference, we continue using a 5\% rate for CutMix because it has no downsides and could make the model more robust to out-of-distribution samples.

We trained the model on a machine with a 16-Core AMD Ryzen 9 7950X CPU, one NVIDIA GeForce RTX 4090, and 126GB of RAM. We implemented the model and training pipeline in TensorFlow~\cite{tensorflow} and Keras~\cite{keras}. The training pipeline handles hundreds of millions of examples in a scalable manner, with dataset sharding and shuffling to ensure that batches are balanced across the different content types. We train the final model for 30 epochs on the training dataset, which took 6 days and 21 hours total on our setup.

Figure~\ref{fig:training-loss-val} shows how the loss and accuracy progress with the number of epochs (computed on the validation split of the dataset). Note how the model achieves a very high validation accuracy already starting from the first few epochs. Figure~\ref{fig:samples-num-vs-f1score} shows how the average F1 score increases with the number of training samples per content type. Note how the average F1 score for binary content types surpasses 99\% with only 10K samples, but that more samples are needed to boost the average F1 score among text content types.

\subsection{Setting Confidence Thresholds}
\label{subsec:threshold-tuning}
A classification threshold is the cut-off point whereby we treat the probability output from our final Dense layer as suitably high-confidence to yield a content-type prediction.
Selecting an optimal threshold involves balancing the trade-offs between precision and recall, which is made more complicated by a multi-class setting. For example, we empirically found that a probability value of 0.80 was robust to assert that a file is \ct{HTML}, whereas the same threshold for \ct{PDF} yielded poor accuracy.

\begin{figure}[t]
    \includegraphics[width=\linewidth]{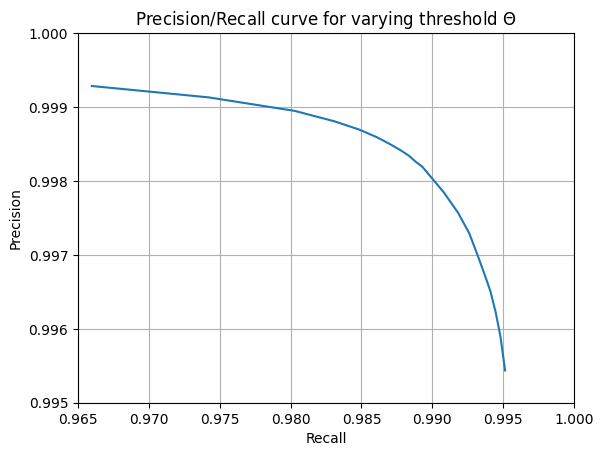}
    \caption{Precision-recall curve for a fixed threshold $\Theta$ applied to all content type predictions. Note that the truncated scale for precision and recall is different.}
    \label{fig:precision-recall-curve}
\end{figure}

We approached this problem by first considering a single threshold for all content types. Figure~\ref{fig:precision-recall-curve} contains the precision-recall curve for this scenario. The curve shows that while there is a trade-off between precision and recall, we can configure the model to obtain a precision and a recall that are both higher than 99.5\%. 

As a further optimization, we instead compute per-content-type thresholds. We determine each threshold by fixing precision at 99\% and then choosing the maximum recall of the remaining thresholds. At inference time, we then select the \textit{argmax} of content type predictions that exceed the type's custom threshold. If the model yields no confident outputs, we output either \ct{txt} or \ct{unknown}, depending on the nature of the content type the model was the most confident with. We discuss the overall performance of this approach shortly. In practice, clients using \magika can set their own confidence thresholds, but by supplying tuned defaults we avoid a pain point encountered by users of other tools~\cite{vscode-guesslang-problems,vscode-guesslang-heuristics}.

\subsection{Performance Optimizations}
Per our design requirements, \magika must also be performant in terms of the time it takes to load the model into memory and yield a prediction. We implement a number of performance optimizations, whereby a client can access our deep learning model either through a command line or API. As part of this, we use OnnxRuntime~\cite{onnxruntime} (instead of Tensorflow and Keras, which we use for training), because it is roughly 15x faster in loading the model (while having a similar model inference time), thus significantly reducing the initialization cost, which is critical for command line use cases.
Likewise, we support batching for clients performing multiple inferences to parallelize processing. Our first CLI prototype has been written in Python, but due to performance reasons, we have also implemented a version in C++ for production use-cases, and a version in Rust as the next version of our command line tool. We have also implemented a prototype based on TensorFlowJS~\cite{tensorflowjs}, which allows one to run \magika entirely within a browser.

\section{Evaluation}
\label{sec:eval}
In order to demonstrate the value of \magika, we benchmark it against existing tools in terms of accuracy and speed.

\subsection{Accuracy}
We report a high-level summary of our benchmark results for \magika in Table~\ref{tab:accuracy-eval-f1score-magika-summary}. The reported accuracy metrics have been computed using the testing dataset, which, following best practices, has not been used for any step involving the creation or tuning of \magika. As a source of further insights, we share per-content type results for a sample of the \ContentTypesNum content types in our benchmark in Table~\ref{tab:accuracy-eval-f1score-magika}.  Full results can be found in our source repository~\cite{magika-anon-website}.  Regardless the context---be it text, binary, or overall content types---we find that \magika outperforms all existing tools. This demonstrates the generalizability of our model architecture and training approach.

\begin{table}[t]
    \centering
    
\begin{tabular}{p{3.1cm}|c|c|c}

    &
    &
    \bf $\Delta$ &
    \bf $\Delta$\\
    
    \textbf{Content Type Metric} & 
    \bf F1 &
    \bf (Supported) & \bf (All) \\
\toprule

Binary & 100\% &  \cellmax4\% &  \cellmax9\% \\ \hline
Text & 99\% &  \cellmax22\% &  \cellmax47\% \\ \hline
Overall & 99\% & \cellmax 12\% & \cellmax 27\% \\
\bottomrule


\end{tabular}
    \medskip
    \caption{Performance of \magika---measured as an F1 score---averaged across all binary, text, and overall content types. To simplify comparison across tools, we calculate the delta ($\Delta$) in F1 score between \magika and the most accurate existing tool per metric.}
    \label{tab:accuracy-eval-f1score-magika-summary}
\end{table}

\mypar{Accuracy gains.} For binary content types, we find that \magika yields a modest F1 gain of 4\% when compared to the best existing tool---\filemime---limited to content types supported by both \magika and \filemime. This F1 gain extends to 9\% if we consider all content types in our benchmark. For text content types, \magika yields even better F1 gains of 22\% over \guesslang, limited to content types supported by both \magika and \guesslang. This extends to 47\% for all content types in our benchmark. This is especially true of \ct{markdown} (45\% gain), \ct{vba} (49\% gain), and \ct{xml} (35\% gain). Likewise, \magika is better than all models at exhibiting uncertainty, with an F1 score of 94\% for \ct{unknown} (synthetic random byte sequences) and 84\% for \ct{txt} (synthetic random text strings).\footnote{In practice, this \ct{txt} includes both TXT files as well as our synthetic examples. This label is only valid if no other text content types apply (\eg \ct{csv}, \ct{html}, \etc.)} 
We also note that \magika has high accuracy (and often shows a double-digit \% gain in F1) even for content types that were shown to be problematic for \guesslang's integration in \vscodeshort, such as \ct{batch}, \ct{csv}, \ct{ini}, \ct{makefile}, \ct{sql}, and \ct{yaml}~\cite{vscode-guesslang-heuristics}.

\mypar{No need for preprocessing.} We find that \magika is able to handle various forms of packing and compression. For example, \ct{doc}, \ct{xls}, \ct{ppt} are all variants of the same Composite Document File binary format. Likewise, \ct{docx}, \ct{xslx}, \ct{pptx}, \ct{apk}, and \ct{jar} are all instances of Zip file formats. Most existing tools deal with these content types by first unpacking the file before applying any signature-based rules. Conversely, \magika requires no pre-processing or domain knowledge to yield accurate predictions.

\begin{table}[t]
    \centering
\begin{minipage}{0.45\columnwidth}
\centering
\begin{tabular}{p{1.4cm}|c|c}
    \textbf{Cont. Type} & 
    \textbf{F1} &
    \textbf{$\Delta$} \\
\toprule
\ct{apk} & 99\% &  \cellmax 16\% \\ \hline
\ct{asm} & 99\% &  \cellmax9\% \\ \hline
\ct{asp} & 99\% &  \cellmax19\% \\ \hline
\ct{batch} & 97\% &  \cellmax32\% \\ \hline
\ct{bmp} & 100\% & 0\% \\ \hline
\ct{c} & 99\% &  \cellmax8\% \\ \hline
\ct{cs} & 100\% &  \cellmax4\% \\ \hline
\ct{css} & 99\% &  \cellmax15\% \\ \hline
\ct{csv} & 99\% &  \cellmax24\% \\ \hline
\ct{dmg} & 100\% &  \cellmax23\% \\ \hline
\ct{doc} & 99\% & -1\% \\ \hline
\ct{docx} & 99\% & 0\% \\ \hline
\ct{elf} & 100\% & 0\% \\ \hline
\ct{go} & 100\% &  \cellmax4\% \\ \hline
\ct{html} & 94\% &  \cellmax13\% \\ \hline
\ct{ini} & 98\% &  \cellmax52\% \\ \hline
\ct{jar} & 98\% &  \cellmax25\% \\ \hline
\ct{java} & 99\% &  \cellmax12\% \\ \hline
\ct{javascript} & 99\% &  \cellmax12\% \\ \hline
\ct{jpeg} & 100\% & 0\% \\ \hline
\ct{json} & 99\% &  \cellmax1\% \\ \hline
\ct{latex} & 100\% &  \cellmax5\% \\\hline
\ct{lisp} & 100\% &  \cellmax4\% \\
\bottomrule
\end{tabular}
\end{minipage}
\begin{minipage}{0.45\columnwidth}
\centering
\begin{tabular}{p{1.6cm}|c|c}
    \textbf{Cont. Type} & 
    \textbf{F1} &
    \textbf{$\Delta$} \\
\toprule
\ct{macho} & 100\% & 0\% \\ \hline
\ct{makefile} & 100\% &  \cellmax5\% \\ \hline
\ct{markdown} & 92\% &  \cellmax45\% \\ \hline
\ct{pdf} & 100\% & 0\% \\ \hline
\ct{pebin} & 100\% & 0\% \\ \hline
\ct{pem} & 100\% &  \cellmax14\% \\ \hline
\ct{perl} & 99\% &  \cellmax20\% \\ \hline
\ct{png} & 100\% & 0\% \\ \hline
\ct{powershell} & 99\% &  \cellmax8\% \\ \hline
\ct{python} & 99\% &  \cellmax5\% \\ \hline
\ct{ruby} & 99\% &  \cellmax11\% \\ \hline
\ct{rust} & 100\% &  \cellmax3\% \\ \hline
\ct{scala} & 100\% &  \cellmax8\% \\ \hline
\ct{shell} & 98\% &  \cellmax9\% \\ \hline
\ct{sql} & 99\% &  \cellmax18\% \\ \hline
\ct{txt} & 84\% &  \cellmax67\% \\ \hline
\ct{unknown} & 94\% &  \cellmax19\% \\ \hline
\ct{vba} & 99\% &  \cellmax49\% \\ \hline
\ct{xls} & 99\% & 0\% \\ \hline
\ct{xlsx} & 99\% & -1\% \\ \hline
\ct{xml} & 99\% &  \cellmax35\% \\ \hline
\ct{yaml} & 99\% &  \cellmax25\% \\\hline
\ct{zip} & 99\% &  \cellmax21\% \\
\bottomrule
\end{tabular}
\end{minipage}
    \medskip
    \caption{Performance of \magika---measured as an F1 score---per content type for a sample of content types in our benchmark. To simplify comparison across tools, we calculate the delta ($\Delta$) in F1 score between \magika and the most accurate existing tool per content type.}
    \label{tab:accuracy-eval-f1score-magika}
\end{table}

\begin{figure}[t]
    \centering
    \includegraphics[width=\columnwidth]{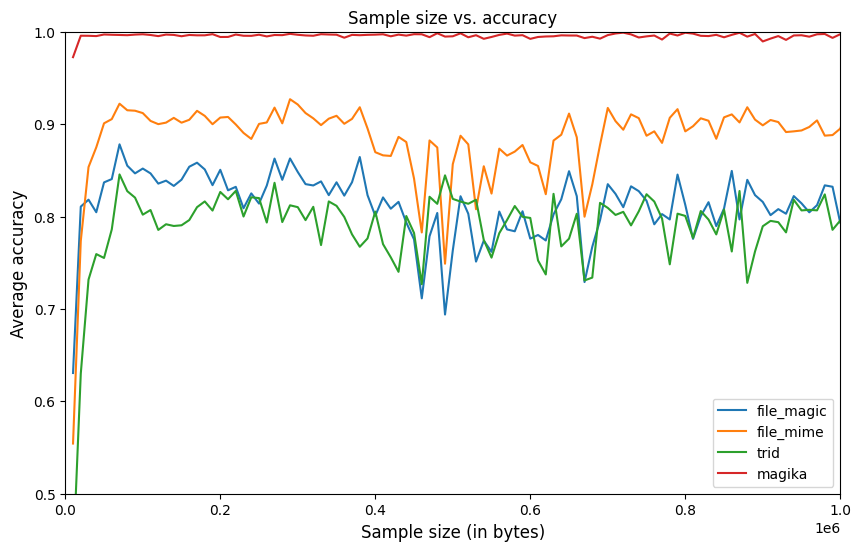}
    \caption{Average accuracy vs. samples size for \magika and existing tools (for simplicity, the figure only shows tools with an average accuracy higher than 50\%).}
    \label{fig:acc-vs-samples-sizes}
\end{figure}

\mypar{Accuracy vs. sample size.} Figure~\ref{fig:acc-vs-samples-sizes} shows how the accuracy of \magika and existing tools change depending on the samples size. The results show how \magika achieves a stable 99\%+ accuracy for most samples, with a slight degrade in accuracy for very small samples of less than $\sim$100 bytes; upon manual inspection, we find that this loss of accuracy is due to \magika outputting a generic content type (such as \ct{TXT} or \ct{UNKNOWN}) due to lack of strong confidence. It is also interesting to see how the average accuracy of existing tools is also somewhat stable across samples size, but with much higher variance.

\mypar{Limitations.} Our evaluation is affected by a few limitations. First, our dataset is balanced across content types: this is standard practice (especially when there is no \textit{one} reference real-world distribution), as it allows one to determine how each content type is supported regardless of whether they are rare in practice. This makes the results more transparent and easier to inspect (e.g., when using an imbalanced dataset, a model with poor support for a rare content type would still have a very high average accuracy). However, we acknowledge that the reported accuracy metrics are not directly applicable, but given that \magika is open-source, a user with a very unique setting can tune and evaluate \magika to its needs, for example by mapping our results to her distribution---this is the standard way to adapt models to an imbalanced environment.

The second limitation is that, as it is common with works based on deep learning on large datasets, we did not perform cross-validation, as it is computationally prohibitively expensive (as reported, a single training run takes about one week). However, cross-validation is deemed as critical only when dealing with very small datasets~\cite{alzubaidi2023survey}, in which a specific ``pick'' of the testing dataset may be accidentally very biased; this is not a concern in our situation, in which the testing set alone has more than 1M samples. We also performed an additional experiment in which we split our testing dataset in 10 random smaller sets: the average accuracy is 99.19\% $\pm$ 0.03\%, showing very low variance, which supports the robustness of our evaluation.

\subsection{Speed}

Per our design requirements, the time it takes to infer a content type is equally important to accuracy. There are two relevant dimensions to processing latency: initialization (\eg loading a signature database or model) and inference (\eg via matching signatures or running a model). For \magika we can measure the breakdown between these two aspects directly, but it is challenging for the other tools. Thus, we measure the breakdown indirectly, by measuring the time required to evaluate one sample at a time, and to evaluate multiple samples all at once.

We evaluate all of the existing tools and \magika as follows. We select a fixed subset of our benchmark dataset that consists of \PerfEvalDatasetSamplesNum samples: ten for each of the \ContentTypesNum content types.\footnote{We selected the first ten samples for each content type. However, one of the \ct{7z} samples causes \exif \texttt{12.65} to hang indefinitely: thus, we excluded this sample and replaced it with another one.} This avoids any bias in the event a tool is faster for one content type versus another (\eg due to requiring decompression or more complex signatures). As a first experiment, we send a single request to each tool for each sample in our evaluation. As a second experiment, we pass all samples as a single request to each tool (\eg we provide each sample's path as arguments to the same command line invocation). In both cases, we limit the number of available CPUs to one (using the \texttt{taskset}~\cite{taskset} utility). We estimate the overall running time for both modes using \texttt{hyperfine}~\cite{hyperfine}, executing each experiment ten times and averaging the results (with three warm up rounds, to minimize the impact of external factors, such as caches). Our environment consists of an isolated docker container within an idle virtual machine (an \texttt{e2-highmem-8} instance on Google Cloud, with 8x AMD Rome CPUs and 64GB of RAM).

\begin{table}[t]
    \centering
    
\begin{tabular}{l|c|c}
    \textbf{Tool name} & \textbf{One sample} & \textbf{All samples}\\ 
                       & \textbf{per request (ms)} & \textbf{in one request (ms)}\\ 
\toprule

\textbf{\file} & 5.36 & 0.75 \\ \hline 
\textbf{\filemime} & 5.30 & 0.67\\ \hline 
\textbf{\exif} & 102.56 & 6.36 \\ \hline 
\textbf{\trid} & 137.08 & 51.57 \\ \hline 
\textbf{\guesslang} & 958.61 & 307.66 \\ \hline 
\textbf{\magika} & 86.73 & 5.77 \\ 

\bottomrule

\end{tabular}

    \medskip
    \caption{Average execution speed of \magika and existing tools when scanning \PerfEvalDatasetSamplesNum samples (\ContentTypesNum content types, 10 samples per type). The first column captures the initialization and inference cost for content-type detection on a single sample (averaged across all samples). The second column captures the amortized initialization and inference costs of content-type detection for all samples in a single request, normalized to the total number of samples.}
    \label{tab:performance-eval}
\end{table}

Table~\ref{tab:performance-eval} shows the results.
We find that \file is the fastest tool (with or without the MIME flag), followed by \magika,  \exif, \trid, and \guesslang. Apart from \file, every tool has a non-negligible initialization cost on the order of 100ms or more. For clients scanning files in bulk, the amortized processing time of \magika drops to 5.77ms, better than all existing tools apart from \file. The significant performance gain of \magika over \guesslang (the only other tool to use a model) is likely due to the latter's reliance on TensorFlowJS~\cite{tensorflowjs}, whereas \magika uses OnnxRuntime. We note that our implementation of \magika allows clients to take advantage of multiple CPUs without having to wrap requests via parallelization, which is not true of existing tools. For example, our client achieves an average of 1.39ms per sample when running on 8 CPUs, while the amortized processing time of the other tools remains the same. While many other optimizations are feasible, our results demonstrate that \magika can quickly infer content types even when executing a neural network exclusively via a CPU.

\section{Real-world Adoption}\label{sec:real-world-deployment}
We recently released the \magika model and a command line wrapper as open source under an Apache 2 license~\cite{magika-github}. We share information on its reception as well as real-world deployments that now use \magika.
 
\mypar{GitHub release.} We reached 4K stars on GitHub in less than a week and currently total more than 7.7K stars. \magika was featured in GitHub trending projects and we have received more than 100 open issues and pull requests from external contributors. Likewise, a number of developers have reached out to integrate \magika into their workflows, such as for validating datasets, or for enhancing security malware analysis pipelines (\eg AssemblyLine~\cite{assemblyline-github}). \magika's Python package currently averages over 3K downloads per day.

\mypar{Email attachment scanner.} We worked with a large email provider, Gmail, to integrate \magika into their attachment scanner that detects and blocks malicious attachments~\cite{magika-gmail-blogpost}. This scanner processes hundreds of billions of files every week, underscoring the computational performance of our approach. \magika allows the email provider to enforce content-type policies on potentially obfuscated files (\eg prohibiting executable binary content types as attachments). It also allows the email provider to route specific samples to anti-virus scanners based on the content type detected, which might otherwise be prohibitively expensive to run on all samples. For example, we have estimated that, on a daily basis, \magika routes to MS Office-specific malware scanners several millions samples that would have otherwise not being scanned when using existing content type detection systems or file extensions. 
\magika has now been running in production for more than six months, with no significant concerns to be reported.

\mypar{VirusTotal file scanner.} The VirusTotal service has recently integrated our work: every submission to VirusTotal is now processed with \magika, whose result can be seen in the ``Details'' tab of each submission, alongside other content type detection tools. This can be used by VirusTotal for improved indexing and to optimize which files are sent to the platform's Code Insight functionality~\cite{vt-code-insights}, which employs generative AI to analyze and detect malicious code. For examples, \magika's accurate detection of \ct{powershell} means that VirusTotal can now isolate such files and specialize any generative analysis exclusively to \ct{powershell}.

\mypar{VS Code.} We have reached out to VS Code developers and we have presented our model for consideration as a potential replacement for \guesslang. During our discussions, the developers acknowledged the limitations and challenges encountered with \guesslang, expressing openness to consider alternatives. We subsequently engaged in a dialogue regarding current feature gaps, primarily focusing on the necessity for more fine-grained detection of specific content types (e.g., \ct{C} vs. \ct{C++}, \ct{javascript} vs. \ct{typescript}, \ct{ini} vs. \ct{toml}), which we have been exploring with very promising results.

\section{Discussion}
Our benchmark and integration stories highlight that content-type detection remains an important problem. We discuss future design directions for further improving \magika and potential risks.

\mypar{Handling new content types.}
We acknowledge that the current version of \magika supports a limited number of content types, and that our selected list is not necessarily the most representative in all scenarios.
Moreover, new content types are constantly emerging, requiring the authors of content-type detection tools to continuously maintain and add new signatures. \magika sidesteps the necessity of  understanding the intricacies of different file formats (\eg to write signatures and to ensure they do not collide with previous signatures). Instead, handling new content types with \magika requires two steps: (1) sourcing a sufficiently large number of training samples; and (2) retraining the model with a new Dense output sized to the number of supported content types. We find in practice that \magika can use as few as 10K samples to achieve robust accuracy for binary content types, though text content types require more samples. Sourcing files is also fairly trivial via GitHub and VirusTotal. While re-training a new model can take several days, it is not a concern as the entire process can be automated.

\mypar{Multiple valid content types.} 
\magika currently outputs only a single inferred content type, which is enough for the vast majority of files. However, one could craft so-called \textit{polyglot} files, which are syntactically valid for multiple content types (\eg interpretable as Python, PHP, and bash)~\cite{koch2022toward}. Such polyglots use unspecified holes in the file format specifications, making accurate detection more challenging. More trivially, multiple valid content types might arise due to a file containing code snippets from different programming languages. In practice, users of \magika may be able to examine the top N most likely content types, though we leave evaluating the accuracy of \magika on multi-content-type files to future work.

\mypar{Adversarial risk.}
As with other content-type detection tools, \magika is potentially susceptible to evasion. Here, evasion means tricking a content-type detection tool to infer an incorrect type, but where the intended application can nevertheless process the file's contents correctly. For signature-based rules, this might be as simple as adding whitespace per our earlier example in Figure~\ref{fig:js-example}. For \magika, as the model is open source, attackers might instead add small perturbations to a file's contents to cause our neural network to infer an incorrect content-type. Adding such noise in a programmatic way, while still yielding a file that is interpretable by the intended application may be non-trivial; however, this analysis falls under the area of adversarial machine learning, and significant work is required to make the approach resilient to attackers that specifically attempt to bypass the detection.

\section{Conclusions}
In this work, we presented \magika, a novel AI-powered content-type detection tool. Trained on \DatasetTrainSamplesNumApproxShort samples and \ContentTypesNum canonical content types, we showed how our deep learning architecture can achieve an average F1 score of 99\%. Our approach outperforms all existing content-type detection tools, with a 4\% F1 gain over the best tool for binary content types and a 12\% F1 gain over the best tool for text content types. For bulk inferences, \magika requires 5.77ms to yield a decision per file with just a single CPU, making it suitable for a variety of deployment scenarios. Our tool has already seen adoption in the real world in a number of critical pipelines. To foster adoption and improvements, parts of \magika are already available as open source under an Apache 2 license, and we plan to release the remaining material in the near future~\cite{magika-github}.

\bibliographystyle{IEEEtran}
\bibliography{bibliography}

\begin{thebibliography}{10}
\bibitem{magika-anon-website}
``{\magika Anonymized Repository: Please see ``Artifact Availability'' section
  of our HotCrp submission for an anonymized version of our repository and
  additional material.}''
\end{thebibliography}

\end{document}